\documentclass[prd,aps,nofootinbib,floatfix,superscriptaddress]{revtex4}
\usepackage{graphicx}
\usepackage{epsfig}
\usepackage{rotating}
\usepackage{amssymb}
\usepackage{subfigure}
\usepackage{dsfont}
\usepackage{psfrag}
\usepackage{amsmath,euscript,array,mathrsfs}

\newcommand{\SP}[1]{\begin{equation}\begin{split} #1
\end{split}\end{equation}}
\newcommand{\beq}{\begin{equation}}
\newcommand{\eeq}{\end{equation}}
\newcommand{\beqs}{\begin{eqnarray}}
\newcommand{\eeqs}{\end{eqnarray}}

\begin{document}
\title{Light scalar from deformations of the Klebanov-Strassler background}

\author{Daniel Elander}
\affiliation{Department of Physics and Astronomy, Purdue University, \\
525 Northwestern Avenue, West Lafayette, IN 47907-2036, USA}

\date{\today}


\begin{abstract}
We study deformations of the Klebanov-Strassler background parametrized by the size of a dim-6 VEV. In the UV, these solutions describe the usual duality cascade of Klebanov-Strassler; however below the scale $\rho_*$ set by the dim-6 VEV they exhibit hyperscaling violation over a range of the radial coordinate. Focusing on the spectrum of scalar glueballs, we find a parametrically light state, the mass of which is suppressed by $\rho_*$, becoming massless in the limit of $\rho_* \rightarrow \infty$. Along the way, we clarify the choice of IR and UV boundary conditions for the fluctuations in the bulk, and find agreement with previous calculations for the spectrum of Klebanov-Strassler.
\end{abstract}

\maketitle

\tableofcontents

\section{Introduction}

Gauge-gravity duality allows for the analytical study of strongly coupled quantum field theories. Since its original inception \cite{AdSCFT} which relates $\mathcal N = 4$ supersymmetric Yang-Mills to Type IIB string theory on $AdS_5 \times S^5$, it has been extended to numerous examples, including cases where conformal symmetry is broken and the field theory confines \cite{Witten,KS,MN}. It is interesting to consider field theories with one or more characteristic scales in addition to that of confinement, thus allowing for more complex dynamics.

One possible application of such multi-scale dynamics is within the context of Technicolor \cite{TC,reviewsTC}, the idea that electroweak symmetry is broken by the dynamics of a strongly coupled field theory. In particular, Walking Technicolor models \cite{WTC,Yamawaki} posit the existence of an energy regime, above the confinement scale, in which the theory is nearly conformal as well as strongly coupled. An attractive feature of such models is that they could potentially contain a light scalar in the spectrum due to the spontaneous breaking of approximate scale invariance \cite{Yamawaki,dilaton}, and that this so-called techni-dilaton would couple to the Standard Model in a similar way as the Higgs \cite{dilatonpheno,dilaton4,dilaton5D,dilatonnew}. Given that a boson with mass 126~GeV has been observed at the LHC \cite{ATLAS,CMS}, it is of great interest to find specific examples of strongly coupled field theories with walking dynamics, and to study whether a composite light dilaton exists in their spectra. This has been the subject of a number of studies within the context of holographic models from string theory \cite{NPP,ENP1,ENP2,MultiScaleKS,stringWTC,stringWTC2}, as well as studies on the lattice \cite{lattice}. While an elementary Higgs certainly is consistent with current experimental data, it is still a possibility that the newly observed particle is a techni-dilaton \cite{dilatonandpheno}.

A string dual of a theory with walking dynamics was found in \cite{NPP} by considering a system of wrapped D5-branes, in the sense that a suitably defined 4d gauge coupling \cite{gaugecoupling} stays nearly constant in an intermediate energy region. The length of this walking region is related to the size of a dim-6 VEV. Below the scale given by the VEV, the metric is hyperscaling violating with exponent $\theta=4$. In the case where the UV is given by the asymptotics of Maldacena-Nunez (MN) \cite{MN}, the spectrum of scalar glueballs can be computed and contains a light state \cite{ENP1,ENP2}, the mass of which is suppressed by the length of the walking region, thus suggesting that it is a dilaton.

There exists a class of solutions that interpolates between Klebanov-Strassler (KS) and MN \cite{Butti}. The relevant parameter is the size of a dim-2 VEV, which is equal to zero for KS, and which for non-zero values takes the theory out on the baryonic branch of KS. Since we know that MN can be deformed by the dim-6 VEV linked to walking dynamics, it is natural to ask whether this also is possible on the baryonic branch of KS. Such solutions were found in \cite{MultiScaleKS} by applying a solution-generating technique \cite{rotation} to the previously mentioned walking solutions of the D5 system of \cite{NPP}. Building on the work of \cite{rotation,quivers}, one can understand the field theory of the new solutions that are generated as follows. In the UV, the duality cascade of KS continues down to a certain scale at which the theory is Higgsed to a single gauge group. Below this scale, the dynamics become the same as that of the original wrapped-D5 solutions from which the new solutions were generated.

In this paper, we consider the case where the dim-2 VEV is turned off, while the dim-6 VEV is turned on. This means that the UV is that of KS deformed by the dim-6 VEV, while below the scale set by this VEV the metric becomes hyperscaling violating, again with the exponent $\theta=4$. This solution was originally given in \cite{MultiScaleKS}, and similar deformations have been considered previously in \cite{Zayas} (see also \cite{Faedo}). It has the advantage of being simpler than the more general walking solutions on the baryonic branch of KS, while at the same time having a UV that is better behaved than that of the deformations of MN.\footnote{A consequence of the UV asymptotics being that of MN is that the spectrum contains a cut, leading to an infinite tower of states converging on $m=1$ and a continuum above $m>1$ ($m$ is the mass given in suitably defined units).}

In order to compute the spectrum of scalar glueballs, we make use of the gauge invariant formalism developed in \cite{gaugeinvariant} and generalized in \cite{ESQCD}. Given a consistent truncation to a non-linear sigma model of scalars coupled to five-dimensional gravity, this formalism allows one to systematically study the fluctuations of the metric as well as the scalars in a given background. Suitable boundary conditions \cite{EPBCs} are applied at IR and UV cutoffs which are eventually dispensed with by taking the appropriate limits. In particular, we impose Dirichlet boundary conditions on the fluctuations of the scalar fields in the IR and UV. We find that in addition to a series of towers of states, the spectrum contains a light scalar, the mass of which is suppressed by the size of the dim-6 VEV. This is the main result of this paper.

All the backgrounds mentioned that have the dim-6 VEV turned on suffer from a singularity in the IR. This singularity is of a mild type, in the sense that while (the ten-dimensional) $R_{\mu\nu\rho\sigma}^2$ blows up, $R_{\mu\nu}^2$ and $R$ stay finite. Furthermore, in the cases that the Wilson loop has been studied, it is well-behaved \cite{NPR,MultiScaleKS}. Still, the singular behavior is cause for some concern, in particular with regard to the form of the boundary conditions imposed on the fluctuations in the IR. We show that with our prescription the spectrum converges as the IR cutoff is taken closer and closer to the singularity.

The structure of this paper is as follows. In Section~\ref{sec:formalism}, we review the gauge invariant formalism for the holographic computation of glueball spectra. In Section~\ref{sec:PT5d}, we review the Papadopoulos-Tseytlin ansatz and the relevant 5D truncation. In section~\ref{sec:backgrounds}, we present the backgrounds that are the topic of this paper, and find that for large dim-6 VEV there exists a good analytical approximation. Section~\ref{sec:spectra} contains the numerical study of the scalar glueball spectrum, first of KS as a consistency check, and later of the deformations thereof by the dim-6 VEV. In the final Section~\ref{sec:conclusions}, we conclude with a discussion of the implications of our results as well as suggestions for future directions of research.

\section{Formalism}\label{sec:formalism}

In this section, we review and summarize the gauge invariant formalism for studying linearized fluctuations in non-linear sigma models consisting of a number of scalars coupled to gravity (for further details see \cite{gaugeinvariant,EPBCs}). We focus on the scalar sector of the fluctuations, and write down a particularly simple version Eq.~\eqref{eq:BCs3} of their boundary conditions in the IR and UV.

\subsection{Action and backgrounds}

We start with a $d+1$-dimensional non-linear sigma model consisting of $n$ scalars $\Phi^a$ ($a=1,\cdots,n$) coupled to gravity, whose action is given by
\SP{\label{eq:action}
	S =& \int d^dx \int_{r_1}^{r_2} dr \sqrt{-g} \left( \frac{R}{4} - \frac{1}{2} g^{MN} G_{ab}(\Phi) \partial_M \Phi^a \partial_N \Phi^b -V(\Phi) \right) \\& - \int d^dx \sqrt{-\tilde g} \left( \frac{K}{2} + \lambda_1(\Phi) \right) \Bigg|_{r_1} + \int d^dx \sqrt{-\tilde g} \left( \frac{K}{2} + \lambda_2(\Phi) \right) \Bigg|_{r_2}.
}
In the bulk part of the action, $g_{MN}$ is the $d+1$-dimensional metric, $G_{ab}(\Phi)$ is the non-linear sigma model metric, and $V(\Phi)$ is the potential for the scalars. The parts of the action localized at the boundaries $r_i$ ($i=1,2$) contain the induced metric $\tilde g_{\mu\nu}$, the extrinsic curvature $K$ that appears in the Gibbons-Hawking terms, and boundary potentials $\lambda_i(\Phi$) for the scalar fields that end up determining the boundary conditions for the background and fluctuations around it.

We are interested in backgrounds that only depend on the radial coordinate $r$, and for which the metric has the form of a domain wall
\SP{
	ds_{d+1}^2 = dr^2 + e^{2A(r)} dx_{1,d-1}^2.
}
These satisfy the equations of motion
\SP{
	\bar \Phi''^a +d A' \bar \Phi'^a + \mathcal G^a_{\ bc} \bar \Phi'^b \bar \Phi'^c - V^a &= 0, \\
	d(d-1) A'^2 - 2 \bar \Phi'^a \bar \Phi'_a + 4V &= 0,
}
where $\mathcal G_{abc} = \frac{1}{2} \left( \partial_b G_{ca} +\partial_c G_{ab} - \partial_a G_{bc} \right)$, $V^a = G^{ab} V_b = G^{ab} \frac{\partial V}{\partial \Phi^b}$, $\bar \Phi'_a = G_{ab} \bar \Phi'^b$, and prime denotes differentiation with respect to $r$. The second equation is the Hamiltonian constraint.

In certain cases, there exists a superpotential $W$ such that
\SP{
	V = \frac{1}{2} W^a W_a - \frac{d}{d-1} W^2,
}
and solutions can then be found by solving the first order equations
\SP{\label{eq:eomsW}
	\bar \Phi'^a &= W^a, \\
	A' &= -\frac{2}{d-1} W.
}

\subsection{Equations of motion for fluctuations}

Next, we study fluctuations $\{ \varphi^a, \nu, \nu^\mu, {h^{TT}}^\mu_{\ \nu}, h, H, \epsilon^\mu \}$  around the background:
\SP{
	\Phi^a &= \bar \Phi^a + \varphi^a, \\
	g_{MN} &=
	\left(
	\begin{array}{ll}
	 	\tilde g_{\mu\nu} & \nu_\nu \\
	 	\nu_\mu & 1 + 2\nu + \nu_\sigma \nu^\sigma
	\end{array}
\right), \\
	\tilde g_{\mu\nu} &= e^{2A} (\eta_{\mu\nu} + h_{\mu\nu}), \\
	h^\mu_{\ \nu} &= {h^{TT}}^\mu_{\ \nu} + \partial^\mu \epsilon_\nu + \partial_\nu \epsilon^\mu + \frac{\partial^\mu \partial_\nu}{\Box} H + \frac{1}{d-1} \delta^\mu_{\ \nu} h,
}
where ${h^{TT}}^\mu_{\ \nu} $ is transverse and traceless, $\epsilon^\mu$ is transverse, $\Box = \eta^{\mu\nu}\partial_\mu \partial_\nu$, and $d$-dimensional indices $\mu$, $\nu$ are raised and lowered by the boundary metric $\eta$. Expanding the equations of motion to linear order, and introducing gauge invariant variables
\SP{
	\mathfrak a^a &= \varphi^a - \frac{\bar{\Phi}^{\prime\,a}}{2(d-1) A'} h, \\
	\mathfrak b &= \nu - \frac{\partial_r (h/A')}{2(d-1)}, \\
	\mathfrak c &= e^{-2A} \partial_\mu \nu^\mu - \frac{e^{-2A} \Box h}{2(d-1) A'} - \frac{1}{2} \partial_r H, \\
	\mathfrak d^{\mu} &= e^{-2A} \Pi^\mu_{\,\,\,\nu} \nu^\nu - \partial_r \epsilon^\mu, \\
	\mathfrak e^{\mu}_{\ \nu} &= {h^{TT}}^\mu_{\ \nu},
}
the spin-1 ($\mathfrak d^\mu$) and spin-2 ($\mathfrak e^{\mu}_{\ \nu} $) sectors decouple, and furthermore $\mathfrak b$ and $\mathfrak c$ can be solved for algebraically in terms of $\mathfrak a^a$, so that the equations of motion for the scalar sector reduce to $n$ coupled second order differential equations for $\mathfrak a^a$ given by
\SP{\label{eq:fluceoms}
	\Big[ \mathcal D_r^2 + d A' \mathcal D_r + e^{-2A} \Box \Big] \mathfrak{a}^a - \Big[ V^a_{\ |c} - \mathcal{R}^a_{\ bcd} \bar \Phi'^b \bar \Phi'^d + \frac{4 (\bar \Phi'^a V_c + V^a \bar \Phi'_c )}{(d-1) A'} + \frac{16 V \bar \Phi'^a \bar \Phi'_c}{(d-1)^2 A'^2} \Big] \mathfrak{a}^c = 0.
}
Here, $V^a_{\ |b} = \frac{\partial V^a}{\partial \Phi^b} + \mathcal G^a_{\ bc} V^c$, $\mathcal R^a_{\ bcd} = \partial_c \mathcal G^a_{\ bd} - \partial_d \mathcal G^a_{\ bc} + \mathcal G^a_{\ ce} \mathcal G^e_{\ bd} - \mathcal G^a_{\ de} \mathcal G^e_{\ bc}$ is the Riemann tensor corresponding to the sigma model metric, while the background covariant derivative is defined as $\mathcal D_r \mathfrak a^a = \partial_r \mathfrak a^a + \mathcal G^a_{\ bc} \bar \Phi'^b \mathfrak a^c$.

Explicitly expanding out the background covariant derivatives $\mathcal D_r$ occurring in Eq.~\eqref{eq:fluceoms}, we obtain
\SP{
\label{eq:flucdiffnoncovariant}
	& \partial_r^2 \mathfrak{a}^a + \left( 2 \mathcal{G}^a_{\ bc} \bar \Phi'^c + d A' \delta^a_{\ b} \right) \partial_r \mathfrak{a}^b + e^{-2A} \Box \mathfrak{a}^a - \\ & \Bigg[ \partial_b V^a  - \partial_b \mathcal{G}^a_{\ cd} \bar \Phi'^c \bar \Phi'^d + \frac{4 (\bar \Phi'^a V_b + V^a \bar \Phi'_b)}{(d-1)A'} + \frac{16 V \bar \Phi'^a \bar \Phi'_b}{(d-1)^2 A'^2} \Bigg] \mathfrak{a}^b = 0,
}
which is sometimes more convenient to use.

\subsection{Boundary conditions for fluctuations}
Boundary conditions for the fields are obtained by varying the action Eq.~\eqref{eq:action} with respect to the metric and the scalar fields, and focusing on the boundary contributions. Expanding the boundary potentials $\lambda_i(\Phi)$ as
\SP{
	\lambda_i &= \left( -\frac{d-1}{2} A' + \bar \Phi'_a \varphi^a + \frac{1}{2} \lambda_{ia|b} \varphi^a \varphi^b \right) \Big|_{r_i} ,
}
where the first and second terms are chosen for consistency at the level of the background itself, and the choice of $\lambda_{ia|b}$ can be thought of as parametrizing the form of the boundary conditions for the fluctuations. In particular, it enters the boundary conditions for $\mathfrak a^a$ as follows:
\SP{
\label{eq:BCs1}
	&\left[ \delta^a_{\ b} + \frac{e^{2A}}{\Box} \left( V^a - d A' \bar \Phi'^a - {\lambda_i}^a_{\ |c} \bar \Phi'^c \right) \frac{2 \bar \Phi'_b}{(d-1) A'} \right] \mathcal D_r \mathfrak a^b \Big|_{r_i} = \\&  \left[ {\lambda_i}^a_{\ |b} + \frac{2 \bar \Phi'^a \bar \Phi'_b}{(d-1) A'} + \frac{e^{2A}}{\Box} \frac{2}{(d-1) A'} \left( V^a - d A' \bar \Phi'^a - {\lambda_i}^a_{\ |c} \bar \Phi'^c \right) \left( \frac{4 V \bar \Phi'_b}{(d-1) A'} + V_b \right) \right] \mathfrak a^b \Big|_{r_i}.
}

The limit of ${\lambda_i}^a_{\ |b} \rightarrow \pm \infty$ (with the sign depending on $i=1,2$), corresponds to Dirichlet boundary conditions for the scalar fluctuations $\varphi^a$
\beq
	\varphi^a \Big|_{r_i} = 0.
\eeq
This is the choice that we will make in the following. In terms of the gauge invariant variables $\mathfrak a^a$, it translates to
\SP{
\label{eq:BCs2}
	-\frac{e^{2A}}{\Box} \frac{2 \bar \Phi'^a}{(d-1) A'} \left[ \bar \Phi'_b \mathcal D_r  - \frac{4 V \bar \Phi'_b}{(d-1) A'} - V_b \right] \mathfrak a^b \Big|_{r_i} = \mathfrak a^a \Big|_{r_i}.
}
We can simplify Eq.~\eqref{eq:BCs2} further by noticing that the matrix multiplying $\mathfrak a^a$ is of the form $M^a_{\ b} = \delta^a_{\ b} + X^a Y_b$, whose inverse is given by $(M^{-1})^a_{\ b} = \delta^a_{\ b} - \frac{X^a Y_b}{1+X^c Y_c}$, and hence
\SP{
\label{eq:BCs3}
	e^{2A} \Box^{-1}  \bar \Phi'^a \bar \Phi'_b D_r \mathfrak a^b \Big|_{r_i} + \frac{(d-1)A'}{2}  \left( 1 + e^{2A} \Box^{-1} \frac{A'}{2} \partial_r \left( \frac{A''}{A'^2} \right) \right) \mathfrak a^a \Big|_{r_i} = 0,
}
where we have also used that
\SP{
	\frac{A'}{2} \partial_r \left( \frac{A''}{A'^2} \right) = -\frac{2}{d-1} \frac{\bar \Phi'^a}{A'} \left( V_a + \frac{4V}{d-1} \frac{\bar \Phi'_a}{A'} \right).
}

\section{Papadopoulos-Tseytlin ansatz and 5D truncation}\label{sec:PT5d}

The Type IIB supergravity backgrounds that we will study fall into the Papadopoulos-Tseytlin ansatz \cite{PT} (which in turn is a truncation of \cite{consistentconifold}). Furthermore, they can be discussed in terms of a 5-dimensional sigma model that is a consistent truncation of the original 10-dimensional system, thus lending themselves well to the formalism discussed in the previous section.

\subsection{Papadopoulos-Tseytlin ansatz}

The Papadopoulos-Tseytlin ansatz is given by
\beqs
	ds^2&=&e^{2p-x} ds_5^2 + (e^{x+g} + a^2 e^{x-g}) (e_1^2 + e_2^2) + e^{x-g} \left( e_3^2 + e_4^2 + 2a (e_1 e_3 + e_2 e_4) \right) + e^{-6p-x} e_5^2, \\
	ds_5^2&=&dr^2 + e^{2A} dx_{1,3}^2, \\
	F_3&=& N \left[ - e_5 \wedge \left( e_4 \wedge e_3 + e_2 \wedge e_1 + b (e_4 \wedge e_1 - e_3 \wedge e_2) \right) + dr \wedge \left( \partial_r b ( e_4 \wedge e_2 + e_3 \wedge e_1) \right) \right], \\
	H_3&=&-h_2 e_5 \wedge (e_4 \wedge e_2 + e_3 \wedge e_1 ) + d r \wedge \Big[ \partial_r h_1 (e_4 \wedge e_3 + e_2 \wedge e_1 ) - \\ & & \partial_r h_2 (e_4 \wedge e_1 - e_3 \wedge e_2 ) + \partial_r \chi (-e_4 \wedge e_3 + e_2 \wedge e_1 ) \Big], \\
	F_5&=&\tilde F_5 + \star \tilde F_5, \ \ \tilde F_5 = -{\cal K} e_1 \wedge e_2 \wedge e_3 \wedge e_4 \wedge e_5,
\eeqs
where
\beqs
e_1&=&-\sin\theta \, d\phi\,,\\
e_2&=&d\theta\,,\\
e_3&=&\cos\psi\sin\tilde{\theta} \, d\tilde{\phi}\,-\,\sin\psi \, d\tilde{\theta}\,,\\
e_4&=&\sin\psi\sin\tilde{\theta} \, d\tilde{\phi}\,+\,\cos\psi \, d\tilde{\theta}\,,\\
e_5&=&d\psi+\cos\tilde{\theta} \, d\tilde{\phi}+\cos{\theta} \, d\phi,
\eeqs
and with constraints
\beqs
	\mathcal K & = & M + 2N (h_1 + b h_2)\,,\\
	\partial_r \chi &=& \frac{(e^{2g}+2a^2+e^{-2g}a^4-e^{-2g})\partial 
	h_1+2a(1-e^{-2g}+a^2e^{-2g})\partial_r h_2}{e^{2g}+(1-a^2)^2e^{-2g}+2a^2}.
\eeqs
In addition, there is also the dilaton field $\phi$. The background fields $\{ g, x, p, \phi, a, b, h_1, h_2 \}$ as well as the warp factor $A$ are presumed to depend only on the radial coordinate $r$. $M$ and $N$ are constants associated with the number of D3- and D5-branes, respectively.

\subsection{5D truncation}

There exists a consistent truncation to a 5D sigma model consisting of the scalars $\Phi = \{ g, x, p, \phi, a, b, h_1, h_2 \}$ appearing in the Papadopoulos-Tseytlin ansatz (now promoted to 5D fields) coupled to gravity, and described by the action \cite{gaugeinvariant}
\beqs
	S&=&\int d^5 x \sqrt{-g}\left(\frac{1}{4}R-\frac{1}{2}G_{ab}(\Phi)\partial_M\Phi^a\partial^M\Phi^b-V(\Phi) \right)\,,
\eeqs
with kinetic terms
\beqs
G_{ab}\partial_M\Phi^a\partial_N\Phi^b
&=&
\frac{1}{2}\partial_M g \partial_N g 
\,+\,\partial_M x \partial_N x
\,+\,6\partial_M p \partial_N p
\,+\,\frac{1}{4}\partial_M \phi \partial_N \phi \\
&&+\,\frac{1}{2}e^{-2g}\partial_M a \partial_N a
+\frac{1}{2}N^2e^{\phi-2x}\partial_M b \partial_N b \nonumber\\
\nonumber &&
+\frac{e^{-\phi-2x}}{e^{2g}+2a^2+e^{-2g}(1-a^2)^2}
\Big[ (1+2e^{-2g}a^2)\partial_M h_1 \partial_N h_1 \\ &&
+\frac{1}{2}(e^{2g}+2a^2+e^{-2g}(1+a^2)^2)\partial_M h_2 \partial_N h_2
+2a(e^{-2g}(a^2+1)+1)\partial_M h_1 \partial_N h_2 \Big] \nonumber
\eeqs
and potential
\beqs
V&=&-\frac{1}{2}e^{2p-2x} \left[e^{g}+(1+a^2)e^{-g} \right] + \frac{1}{8}e^{-4p-4x} \left[ e^{2g}+(a^2-1)^2e^{-2g}+2a^2 \right]\nonumber\\
&&\,+\,\frac{1}{4}a^2e^{-2g+8p} + \frac{1}{8}N^2e^{\phi-2x+8p}\left[e^{2g}+e^{-2g}(a^2-2a b +1)^2 +2 (a-b)^2\right]\nonumber\\
&&\,+\,\frac{1}{4}e^{-\phi-2x+8p}h_2^2 + \frac{1}{8}e^{8p-4x} \left[ M+2N(h_1+b h_2) \right]^2\,.\nonumber
\eeqs
Any solution to the equations of motion following from this 5D model can be uplifted to a 10d solution in Type IIB supergravity.

Finally, note that (as long as $N \neq 0$) the change of variables
\SP{
	\tilde h_1 &= \frac{M}{2} + N h_1, \\
	\tilde h_2 &= N h_2, \\
	\tilde \phi &= \phi + 2 \log N,
}
removes all dependence on $M$ and $N$ from the 5D model.

\section{Backgrounds}\label{sec:backgrounds}

In this section, we describe the Type IIB supergravity backgrounds that we will study. They are deformations of Klebanov-Strassler corresponding to a dim-6 VEV.

\subsection{Klebanov-Strassler and deformations}

The backgrounds that we are interested in can be found by further truncating the system as
\SP{
	a =& \tanh(y), \\
	g =& -\log(\cosh(y)),
}
in which case there exists a superpotential
\SP{
	W = \frac{1}{2} \left( e^{4 p-2 x} \left(\tilde h_1+ b \tilde h_2\right) - e^{4 p} \cosh (y) - e^{-2 p-2 x} \right).
}

It is practical to change the radial coordinate from $r$ to $\rho$ defined by
\SP{
	d\rho = \frac{1}{2} e^{4p} dr,
}
after which we can write down solutions to the equations of motion Eq.~\eqref{eq:eomsW} given by
\SP{
	y &= 2 \ \rm arctanh (e^{-2\rho}), \\
	\tilde \phi &= \tilde \phi_0, \\
	b &= \frac{2\rho}{\sinh(2\rho)}, \\
	\tilde h_1 &= e^{\tilde \phi_0} \left(2 \rho  \coth (2 \rho )-1\right) \coth (2 \rho ), \\
	\tilde h_2 &= e^{\tilde \phi_0} \frac{1-2 \rho  \coth (2 \rho )}{\sinh(2 \rho )}, \\
	p &= \frac{1}{6} \log \left(\frac{3}{4} \frac{f_0-4\rho +\sinh(4\rho)}{\sinh^2(2 \rho )} \right)-\frac{x}{3}, \\
	A &= A_0 + \frac{\tilde \phi_0}{2}+ \frac{x + \log \sinh (2 \rho )}{3} ,	
}
where $\tilde \phi_0$, $f_0$, and $A_0$ are integration constants, and $x$ satisfies the differential equation
\SP{\label{eq:eomx}
	\partial_\rho x + 2(\tilde h_1 + b \tilde h_2) e^{-2x} - 2e^{-2x-6p} = 0.
}
We fix the boundary conditions in the UV so that $x$ matches the solution due to Klebanov-Tseytlin \cite{KT}
\SP{
	x_{KT} = \frac{\tilde \phi_0}{2} + \frac{1}{2} \log \left(3 \rho -\frac{3}{8}\right).
}
Essentially, this corresponds to fixing an integration constant so that the divergence of $x$ in the UV is softened. We can write solutions to Eq.~\eqref{eq:eomx} satisfying this boundary condition as
\SP{\label{eq:xint}
	x &= \frac{\tilde \phi_0}{2}-\frac{\log 2}{6} + \frac{1}{3} \log \left(f_0-4 \rho +\sinh (4 \rho )\right)+\frac{1}{2} \log \mathcal I, \\
   	\mathcal I(\rho) &= \int_\rho^\infty d\tilde \rho \, \frac{4 \sqrt[3]{2} (2 \tilde \rho  \coth (2 \tilde \rho )-1) \left(\coth (2 \tilde \rho )-2 \tilde \rho \,
   \text{csch}^2(2 \tilde \rho )\right)}{\left(f_0-4 \tilde \rho +\sinh (4 \tilde \rho
   )\right)^{2/3}}.
}
From the UV expansion (see also Appendix~\ref{sec:UVexpansions})
\SP{
	x_{UV} = \frac{\tilde \phi_0}{2} + \frac{1}{2} \log \left(3 \rho-\frac{3}{8}\right) +
	\frac{2 \left(5 f_0 (40 \rho +1)-4 (5 \rho  (80 \rho
   -31)+8)\right)}{125 (8 \rho -1)} e^{-4 \rho} + \mathcal O(e^{-8\rho}),
   }
we see that there is a characteristic scale $\rho_* \equiv \frac{1}{4} \log f_0$ above which the UV expansion is valid, and below which the solution changes radically.

All these solutions have their end of space in the IR at $\rho = 0$. The background of Klebanov-Strassler corresponds to putting $f_0=0$, in which case the IR geometry is that of the deformed conifold. Non-zero values of $f_0$ correspond to turning on a dim-6 VEV,\footnote{The field theory operators corresponding to the various fields in the bulk can be found in \cite{BCPZ}, and in particular the dim-6 VEV under consideration is for the operator $\rm Tr \, W^2 \bar W^2$.} in which case there is a curvature singularity in the IR ($R_{\mu\nu\rho\sigma}^2$ diverges, while $R_{\mu\nu}^2$ and $R$ stay finite).

There exists in the literature various proposals for determining whether a gravity background can be expected to capture the relevant physics despite having an IR singularity. By considering whether a given background admits finite temperature generalizations, Gubser \cite{Gubser:2000nd} arrived at a criterion in terms of the five-dimensional potential $V$ evaluated on the background, namely that it should be bounded from above in order for a singularity to be 'good'. In \cite{Maldacena:2000mw}, Maldacena and Nunez formulated a criterion in terms of the ten-dimensional metric, the strong version of which states that $g_{tt}$ should not increase as one approaches the singularity (this was motivated by the interpretation of the radial coordinate as corresponding to energy scale). The backgrounds considered in this paper satisfy this latter criterion.

\subsection{Approximate solutions}

For $\rho \ll \rho_*$, we can approximate $\mathcal I$ appearing in Eq.~\eqref{eq:xint} as
\SP{\label{eq:hIR}
	\mathcal I_{approx}^{(1)} = 4 \sqrt[3]{2} e^{-\frac{8 \rho _*}{3}} \left[ \mathcal I_0 -\rho ^2 \coth ^2(2 \rho
   )+\rho  \coth (2 \rho ) + \mathcal O(e^{-4(\rho_*-\rho)}) \right],
}
where $\mathcal I_0$ is an integration constant to be determined in a moment. On the other hand, for $\rho \gg 1$ we can make the approximation
\SP{\label{eq:hpapprox}
	\partial_\rho \mathcal I = \frac{8 (1-2 \rho )e^{-\frac{8 \rho _*}{3}}}{\left(2+e^{4 (\rho - \rho_*)}\right)^{2/3}} + \mathcal O( e^{-4\rho -\frac{8}{3} \rm{max}(\rho,\rho_*) }),
}
which gives
\SP{\label{eq:happrox}
	\mathcal I_{approx}^{(2)} = e^{-8 \rho /3} \Bigg[ \frac{9}{4}  \,
   _3F_2\left(\frac{2}{3},\frac{2}{3},\frac{2}{3};\frac{5}{3},\frac{5}{3};-2
   e^{4 (\rho _*- \rho) }\right)+ 3 (2 \rho -1) \,
   _2F_1\left(\frac{2}{3},\frac{2}{3};\frac{5}{3};-2 e^{4 (\rho _*- \rho)
   }\right)\Bigg].
}

\begin{figure}[t]
\begin{center}
\begin{picture}(300,210)
\put(0,0){\includegraphics[width=300pt]{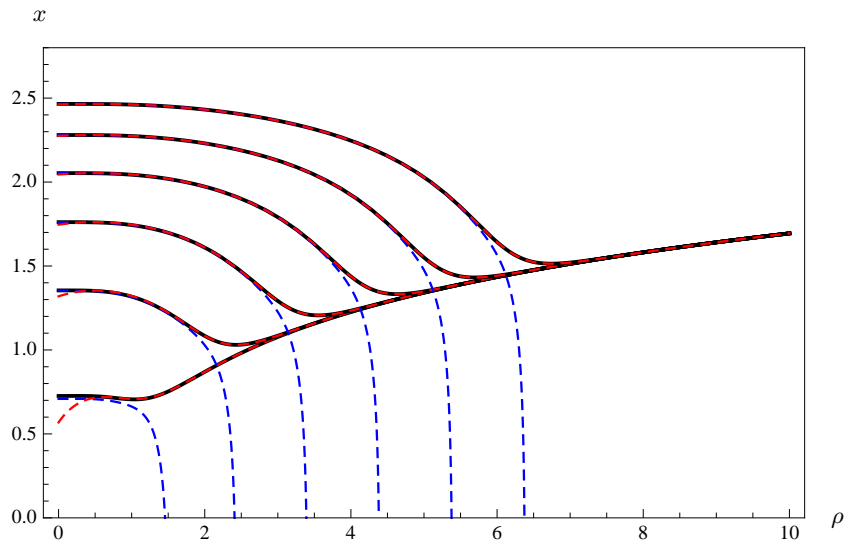}}
\put(310,8){$\rho$}
\put(8,198){$x$}
\end{picture}
\caption{The black lines are $x$ obtained numerically for a few values of $\rho_*$, whereas the dotted blue and red lines are $x$ evaluated using the approximations $\mathcal I_{approx}^{(1)}$ and $\mathcal I_{approx}^{(2)}$, respectively.}
\label{fig:approx}
\end{center}
\end{figure}

For $\rho_* \gg 1$, these two approximations overlap in the region $1 \ll \rho \ll \rho_*$. Matching gives
\SP{
	\mathcal I_0 &= \rho_*^2+\frac{1}{2} \rho _* \Bigg[-2-\gamma +\log (2)-\psi
   ^{(0)}\left(\frac{2}{3}\right)\Bigg] +\frac{1}{64} \Bigg[ \pi ^2+9 \log
   ^2(3)-8 \gamma  (\log (2)-2) \\& -\sqrt{3} \pi  \log (9) +16 \psi
   ^{(0)}\left(\frac{2}{3}\right)+4 \psi ^{(1)}\left(\frac{2}{3}\right)+4 \log
   (2) \left(-4+\log (2)-2 \psi ^{(0)}\left(\frac{2}{3}\right)\right)\Bigg].
}
Using this, let us examine how good the approximation $\mathcal I_{approx}^{(2)}$ is in the region $0 \leq\rho \sim 1$. Expanding for large $\rho_*$, we obtain
\SP{
	\mathcal I_{approx}^{(2)} = \mathcal I_{approx}^{(1)} + 4 \sqrt[3]{2} e^{-\frac{8 \rho _*}{3}} \rho  (\coth (2 \rho )-1) (\rho +\rho 
   \coth (2 \rho )-1) + \mathcal O(e^{-\frac{20\rho_*}{3}}).
}
Since $\mathcal I_{approx}^{(1)}$ is of order $\mathcal O(\rho_*^2 e^{-\frac{8\rho_*}{3}})$, $\mathcal I_{approx}^{(2)}$ as given by Eq.~\eqref{eq:happrox} is actually a good approximation all the way to $\rho =0$ (as long as $\rho_*$ is moderately large), and hence for all values of $\rho$. We illustrate the accuracy of the two approximations $\mathcal I_{approx}^{(1)}$ and $\mathcal I_{approx}^{(2)}$ in Figure~\ref{fig:approx}.

\subsection{Hyperscaling violating region}

A background whose metric transforms as
\SP{
	 ds_{d+1}^2 \rightarrow \lambda^{\frac{2\theta}{d-1}} ds_{d+1}^2
}
under scale transformations is said to be hyperscaling violating with hyperscaling violation exponent $\theta$.

Consider the region $\rho \ll \rho_*$. The metric can be approximated as
\SP{
	ds_5^2 = 4 e^{-8p} d\rho^2 + e^{2A} dx_{1,3}^2 \approx e^{\frac{4\tilde \phi_0}{3}} \Bigg[ \frac{16\ 2^{2/3} \rho _*^{8/3}}{3 \sqrt[3]{3}} e^{\frac{16}{3} (\rho - \rho_*)} d\rho^2 + \rho _*^{2/3} e^{2A_0 + \frac{4\rho}{3}} dx_{1,3}^2 \Bigg],
}
which under the scale transformation
\SP{
	x &\rightarrow \lambda x, \\
	\rho &\rightarrow \rho + \frac{1}{2} \log \lambda,
}
transforms as $	ds_5^2 \rightarrow \lambda^{\frac{8}{3}} ds_5^2$. Hence, in this region the metric is hyperscaling violating with $\theta=4$, which is the same as found for the backgrounds of \cite{NPP,ENP1,ENP2}. Notice the fact that the rescaling of the metric can be compensated by simultaneously transforming $\tilde \phi_0$ as
\SP{
	\tilde \phi_0 &\rightarrow \tilde \phi_0 - 2 \log \lambda.
}

\section{Spectra}\label{sec:spectra}

In this section, we numerically compute the spectrum of scalar glueballs first for Klebanov-Strassler as a warm-up exercise, and then for the deformations thereof. This is done by studying fluctuations of the scalar fields and the metric in the bulk, using the gauge invariant formalism explained in Section~\ref{sec:formalism}. In particular, after imposing the boundary conditions Eq.~\eqref{eq:BCs3} in the IR and UV, there are only certain values of $m^2=\Box$ for which the equations of motion Eq.~\eqref{eq:fluceoms} for the fluctuations can be satisfied. These $m$ give us the spectrum. It is practical to use the so-called midpoint determinant method outlined in \cite{BHMspectrumKS} to study this problem numerically.

In principle, the spectrum depends on the three integration constants $\tilde \phi_0$, $f_0$, and $A_0$. However, $A_0$ simply sets an overall scale, and since we are only interested ratios of masses, it drops out. Furthermore, the spectrum does not depend on $\tilde \phi_0$. The way to see this is that in the 5D sigma model, combinations of $A$, $x$, $p$, $\tilde \phi$, $\tilde h_1$, and $\tilde h_2$ always appear in such a way that $\tilde \phi_0$ can be eliminated from Eq.~\eqref{eq:fluceoms} and Eq.~\eqref{eq:BCs3} after rescaling the fluctuations corresponding to $\tilde h_1$ and $\tilde h_2$ by a factor $e^{\tilde \phi_0}$.\footnote{This was demonstrated for the case $f_0=0$ in \cite{BHMspectrumKS}, but can also be shown to be true for general values of $f_0$.} Therefore, the only interesting variable on which the spectrum depends is $f_0 = e^{4\rho_*}$.

\subsection{Warm-up: Klebanov-Strassler}

The glueball spectrum of Klebanov-Strassler was computed in \cite{BHMspectrumKS} (see also \cite{KrasnitzCaceres,Melnikov}). There, the boundary conditions on the fluctuations were obtained by studying the IR and UV behavior of the fluctuations, picking normalizable solutions in the UV and imposing regularity in the IR. Because of mixing between the fluctuations, this is a complicated problem. On the other hand, the boundary conditions Eq.~\eqref{eq:BCs3} can easily be imposed without any detailed knowledge of the asymptotic  behavior of the fluctuations. For a particularly simple example consisting of a single scalar with quadratic superpotential $W$, it was argued in \cite{ENP2} that these boundary conditions automatically pick the normalizable modes in the UV, and (together with the IR boundary condition) lead to physically sensible results. It is not obvious that this is true in general. Therefore, before we proceed to the more complicated study of the spectrum as a function of $f_0$, we would like to first see whether the boundary conditions Eq.~\eqref{eq:BCs3} lead to the same spectrum as computed in \cite{BHMspectrumKS} for the case $f_0=0$, i.e. the Klebanov-Strassler background.

\begin{figure}[t]
\begin{center}

\begin{picture}(300,210)
\put(0,0){\includegraphics[width=300pt]{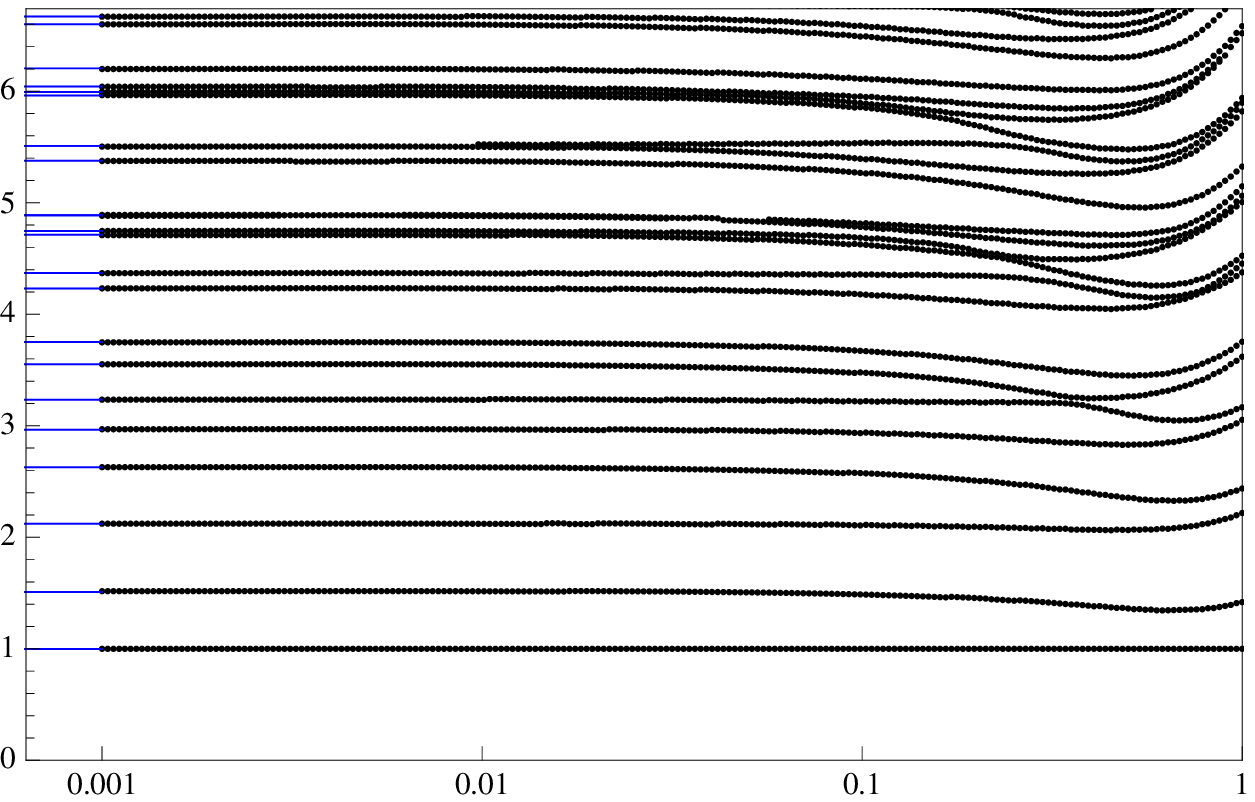}}
\put(310,8){$\rho_{IR}$}
\put(2,205){$m$}
\end{picture}
\caption{Numerical study of the spectrum of scalar glueballs for the Klebanov-Strassler background. The black points show the masses $m$ (normalized to the mass of the first state) as a function of the IR cutoff $\rho_{IR}$, and with constant UV cutoff $\rho_{UV}=11$. The blue lines are the values reported in \cite{BHMspectrumKS} appropriately normalized.}
\label{fig:KSIR}

\begin{picture}(300,230)
\put(0,0){\includegraphics[width=300pt]{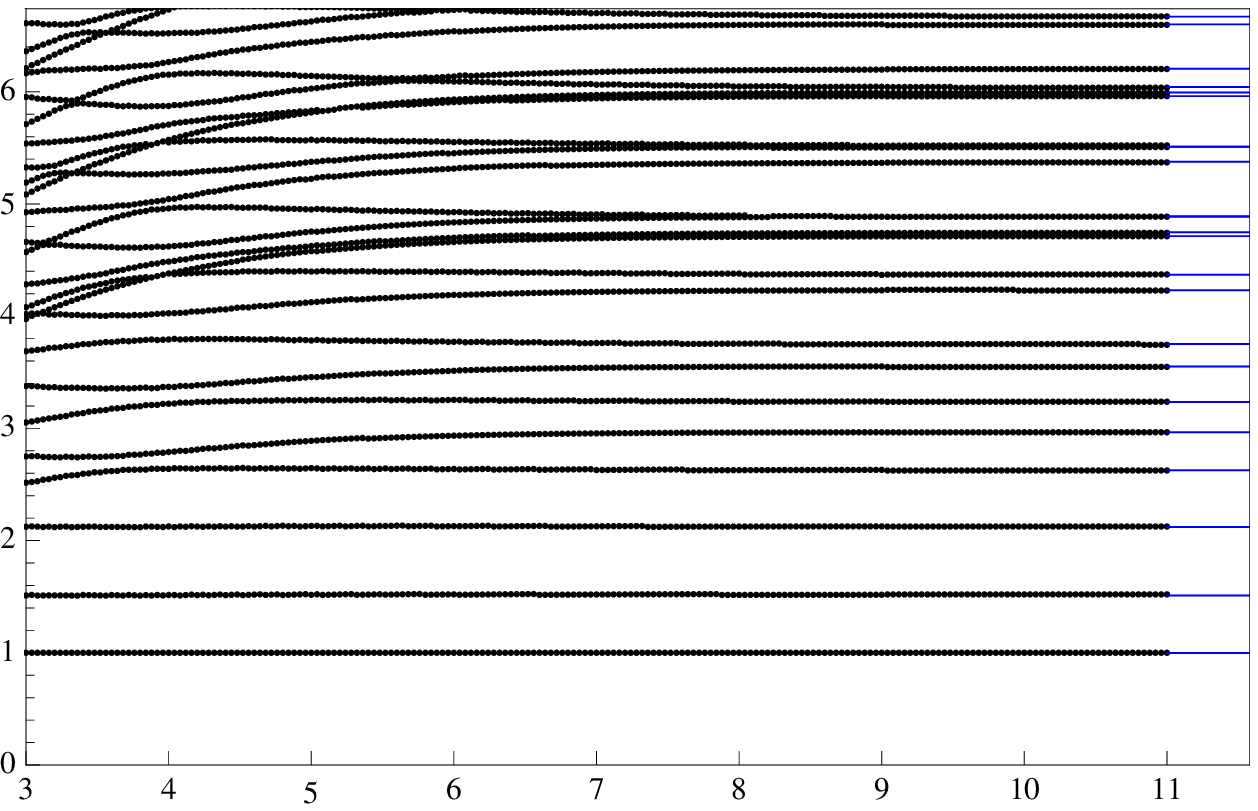}}
\put(310,8){$\rho_{UV}$}
\put(2,205){$m$}
\end{picture}
\caption{Numerical study of the spectrum of scalar glueballs for the Klebanov-Strassler background. The black points show the masses $m$ (normalized to the mass of the first state) as a function of the UV cutoff $\rho_{UV}$, and with constant IR cutoff $\rho_{IR}=0.001$. The blue lines are the values reported in \cite{BHMspectrumKS} appropriately normalized.}
\label{fig:KSUV}

\end{center}
\end{figure}

We specify the boundary conditions Eq.~\eqref{eq:BCs3} at $\rho=\rho_{IR}$ in the IR and at $\rho=\rho_{UV}$ in the UV. We can think of this as introducing IR and UV regulators, which are eventually to be dispensed with by taking the appropriate limits, thus giving us the physical spectrum. In other words, we want to make sure that the calculation of the spectrum converges as $\rho_{IR}$ approaches the end of space at $\rho=0$ while $\rho_{UV}$ approaches infinity. Figure~\ref{fig:KSIR} and Figure~\ref{fig:KSUV} show the dependence of the spectrum as a function of $\rho_{IR}$ and $\rho_{UV}$. For sufficiently small $\rho_{IR}$ and large $\rho_{UV}$, the spectrum becomes the same as that reported in \cite{BHMspectrumKS}. Notice that, as already pointed out in \cite{BHMspectrumKS}, certain states are degenerate (the 13th and 14th states, as well as the 16th and 17th states), and that this is easy to discern in the plots as different states approaching each other asymptotically.

\subsection{Deformations of Klebanov-Strassler}

We now turn to the study of the deformations of Klebanov-Strassler by the dim-6 VEV, in other words non-zero values of $f_0 = e^{4\rho_*}$. Again, we need to make sure that the spectrum converges in the limit $\rho_{IR} \rightarrow 0$ and $\rho_{UV} \rightarrow \infty$. The difference is that there is now a curvature singularity in the IR, which could potentially ruin the convergence of the spectrum in this limit. Fortunately, this is not the case, as shown in Appendix~\ref{IRUVdependence}, where we carry out a study of the dependence of the spectrum as a function of the IR and UV cutoffs.

\begin{figure}[t]
\begin{center}

\begin{picture}(300,210)
\put(0,0){\includegraphics[width=300pt]{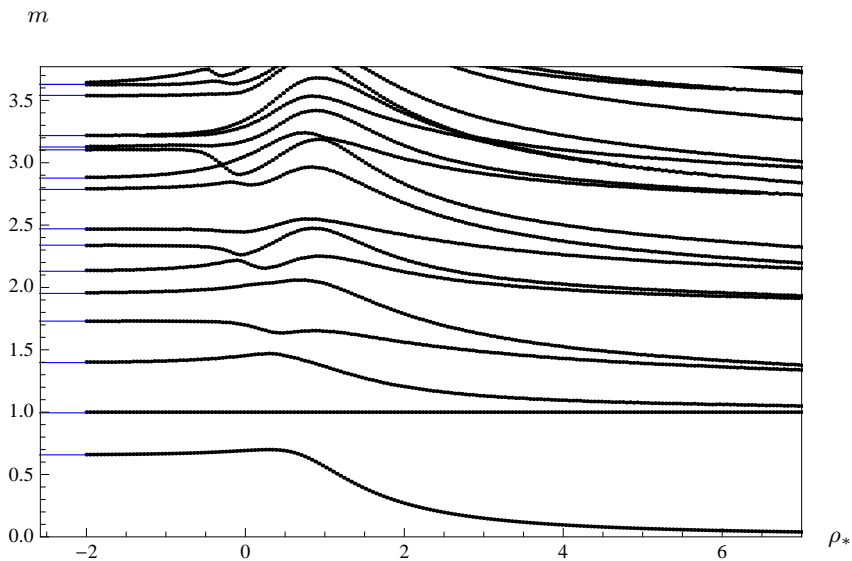}}
\put(310,8){$\rho_*$}
\put(7,205){$m$}
\end{picture}
\caption{Numerical study of the spectrum of scalar glueballs for the deformed Klebanov-Strassler backgrounds. The black points show the masses $m$ (normalized to the mass of the second state) as a function of $\rho_*$, and with constant IR cutoff $\rho_{IR}=0.001$ and UV cutoff $\rho_{UV}=13$. The blue lines show the spectrum for Klebanov-Strassler.}
\label{fig:dKSvaryr0}

\end{center}
\end{figure}

\begin{figure}[t]
\begin{center}

\begin{picture}(300,230)
\put(0,0){\includegraphics[width=300pt]{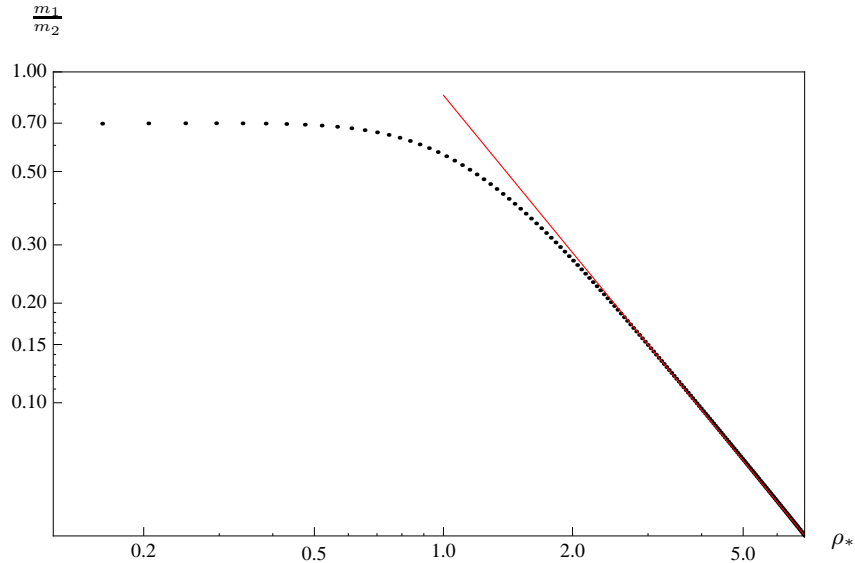}}
\put(310,8){$\rho_*$}
\put(7,205){$\frac{m_1}{m_2}$}
\end{picture}
\caption{Numerical study of the spectrum of scalar glueballs for the deformed Klebanov-Strassler backgrounds. The black points show the ratio of masses of the first and the second lightest states $m_1/m_2$ as a function of $\rho_*$, and with constant IR cutoff $\rho_{IR}=0.001$ and UV cutoff $\rho_{UV}=13$. The red line is proportional to $\rho_*^\alpha$ with $\alpha \approx -1.6$.}
\label{fig:dKSvaryr0LogLog}

\end{center}
\end{figure}

More interestingly, Figure~\ref{fig:dKSvaryr0} shows the dependence of the spectrum on $\rho_*$. As can be seen, for large negative $\rho_*$ the spectrum approaches that of Klebanov-Strassler, while for large positive $\rho_*$, in addition to various towers, there is a light state whose mass is suppressed by $\rho_*$. In particular, as Figure~\ref{fig:dKSvaryr0LogLog} shows, for large $\rho_*$ the mass of this light state falls off with $\rho_*$ according to a power law.

Finally, we would like to make a comment about fine-tuning. For special choices of the boundary conditions for the fluctuations, it is in fact possible to make a light scalar appear in {\it any} theory described by the sigma-model action given in Eq.~\eqref{eq:action}, as pointed out in \cite{EPBCs}. More recently in \cite{EFHMP}, a strongly coupled confining field theory was studied and the parameter $\lambda$ of Eq.~\eqref{eq:BCs1} was varied in order to determine the amount of fine-tuning necessary obtain a light state, something that turned out to be possible only for a very narrow range of $\lambda$. The choice $\lambda \rightarrow \pm \infty$ adopted in this paper is in a sense the most conservative one \cite{EPBCs}: if a light state is present, it is due to the dynamics of the strongly coupled field theory rather than due to fine-tuning of the boundary conditions, and as such it is a robust result.

\section{Conclusions}\label{sec:conclusions}

We first studied the scalar glueball spectrum of Klebanov-Strassler, and found agreement with the previous study carried out in \cite{BHMspectrumKS}. While the boundary conditions used in \cite{BHMspectrumKS} were obtained by studying the IR and UV asymptotics of the fluctuations in the bulk, and imposing regularity and normalizability, respectively, we simply imposed Dirichlet boundary conditions for the scalar fluctuations $\varphi^a |_{IR} = \varphi^a |_{UV} =0$. This choice leads to boundary conditions for the gauge invariant fluctuations $\mathfrak a^a$ given in Eq.~\eqref{eq:BCs3}, and has the advantage that it can easily be implemented without analyzing the IR and UV asymptotics of the fluctuations. The near perfect agreement with the previous study \cite{BHMspectrumKS} suggests that our prescription automatically enforces regularity and normalizability.

The main result of this paper concerns backgrounds which are deformations of Klebanov-Strassler by a dim-6 VEV. The size of this VEV determines the length of a region over which the metric exhibits hyperscaling violation with exponent $\theta=4$. We found that the spectrum of scalar glueballs contains a light state, whose mass is suppressed by the size of the dim-6 VEV. This is analogous to what was found for the walking backgrounds of \cite{ENP1,ENP2} which are deformations of the Maldacena-Nunez background by the same dim-6 VEV. Although it is tempting to interpret the light scalar as being a dilaton, it is presently unclear whether this is actually the case. In order to clarify this issue, one would have to determine precisely which fluctuation in the bulk corresponds to the light state, something which is not immediately clear from the numerical methods used in this paper.

The baryonic branch of Klebanov-Strassler is obtained by turning on a dim-2 VEV. One can view the solutions of Klebanov-Strassler and Maldacena-Nunez as the opposite ends of this baryonic branch. Hence, together with the previous studies of \cite{ENP1,ENP2}, the present study shows that at both these extreme ends, turning on the aforementioned dim-6 VEV gives rise to a light scalar being present in the spectrum. This suggests that such a state should also exist all along the baryonic branch of Klebanov-Strassler if the dim-6 VEV is turned on as in the solutions of \cite{MultiScaleKS}. It would be interesting to see whether this is the case.

\vspace{1.0cm}
\begin{acknowledgments}

We would like to thank Carlos Nunez and Maurizio Piai for valuable discussions. This work is supported by the DOE through grant \protect{DE-SC0007884}.

\end{acknowledgments}

\appendix

\section{UV expansions}\label{sec:UVexpansions}

The UV expansions of the various scalar fields appearing in the background of KS deformed by the dim-6 VEV corresponding to $f_0$ are as follows (the radial coordinate $z$ is defined so that $r=-\frac{3}{2} \log z$):

\beqs
	\tilde \phi &=& \tilde \phi_0, \\
	x &=& \frac{\tilde{\phi }_0}{2} + \frac{1}{2} \log \left(-\frac{3}{8} (12 \log
   (z)+1)\right)+\frac{2 \left(30 \log (z) \left(10 f_0+120 \log
   (z)+31\right)-5 f_0+32\right)}{125 (12 \log (z)+1)}z^6 +\mathcal O\left(z^{12}\right), \\
	p &=& - \frac{\tilde{\phi }_0}{6} - \frac{1}{6} \log \left(-3 \log
   (z)-\frac{1}{4}\right)+\frac{\left(30 \log (z) \left(30 f_0+60
   \log (z)+13\right)+135 f_0+61\right)}{375 (12 \log
   (z)+1)}z^6 +\mathcal O\left(z^{12}\right), \\
   g &=& -2 z^6+\mathcal O\left(z^{18}\right), \\
   a &=& 2 z^3+ \mathcal O\left(z^9\right), \\
   b &=& -6 z^3 \log (z)+\mathcal O\left(z^9\right), \\
   \tilde h_1 &=& -e^{\tilde{\phi }_0} (3 \log (z)+1)-2 e^{\tilde{\phi }_0} (6 \log
   (z)+1)z^6 +\mathcal O\left(z^{12}\right) , \\
   \tilde h_2 &=& 2 e^{\tilde{\phi }_0} (3 \log (z)+1)z^3 +\mathcal O\left(z^{9}\right).
\eeqs

\section{Dependence of the spectrum on the IR and UV regulators}\label{IRUVdependence}

\begin{figure}[t]
\begin{center}

\begin{picture}(300,210)
\put(0,0){\includegraphics[width=300pt]{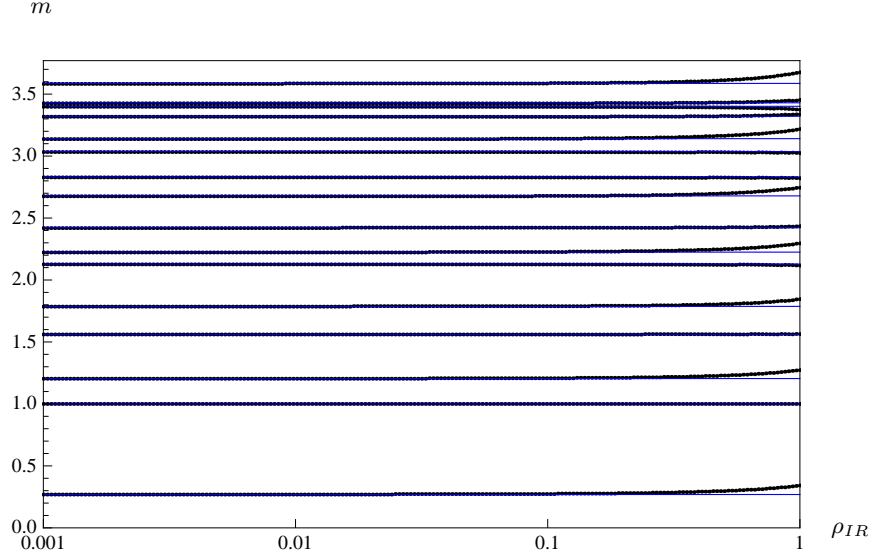}}
\put(310,8){$\rho_{IR}$}
\put(7,205){$m$}
\end{picture}
\caption{Numerical study of the spectrum of scalar glueballs for the deformed Klebanov-Strassler background with $\rho_*=2$. The black points show the masses $m$ (normalized to the mass of the second state) as a function of the IR cutoff $\rho_{IR}$, and with constant UV cutoff $\rho_{UV}=11$. The blue lines show the asymptotic values for small IR cutoff.}
\label{fig:dKSIR}

\end{center}
\end{figure}

\begin{figure}[t]
\begin{center}

\begin{picture}(300,230)
\put(0,0){\includegraphics[width=300pt]{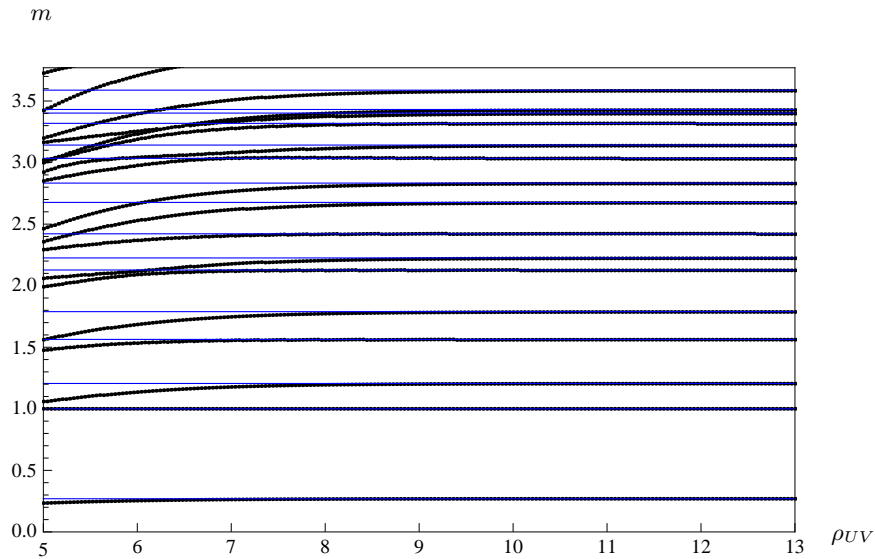}}
\put(310,8){$\rho_{UV}$}
\put(7,205){$m$}
\end{picture}
\caption{Numerical study of the spectrum of scalar glueballs for the deformed Klebanov-Strassler background with $\rho_*=2$. The black points show the masses $m$ (normalized to the mass of the second state) as a function of the UV cutoff $\rho_{UV}$, and with constant IR cutoff $\rho_{IR}=0.001$. The blue lines show the asymptotic values for large UV cutoff.}
\label{fig:dKSUV}

\end{center}
\end{figure}

Figures~\ref{fig:dKSIR} and \ref{fig:dKSUV} show the IR and UV cutoff dependence of the spectrum for the background with $\rho_*=2$. Note that despite the curvature singularity in the IR, the spectrum converges as the IR cutoff is taken towards the end of space.

\end{document}